\documentclass{aastex}          
\usepackage{spr-astr-addons}    

\begin{document}

\title{Generalized second law of thermodynamics in $f(R,T)$ theory of gravity}

\shorttitle{<Short article title>}
\shortauthors{<Momeni et al>}

\author{D. Momeni\altaffilmark{1} , P.H.R.S. Moraes\altaffilmark{2} , R. Myrzakulov \altaffilmark{1}} 

\affil{1:Eurasian International Center for Theoretical Physics and Department of
General \& Theoretical Physics, Eurasian National University, 
Astana 010008, Kazakhstan}
\affil{2:ITA - Instituto Tecnol\'ogico de Aeron\'autica - Departamento de F\'isica, 12228-900, S\~ao Jos\'e dos Campos, S\~ao Paulo, Brazil}
\email{davoodmomeni78@gmail.com}
\email{moraes.phrs@gmail.com} 

\begin{abstract}

We present a study of the generalized second law of thermodynamics in the scope of the $f(R,T)$ theory of gravity, with $R$ and $T$ representing the Ricci scalar and trace of the energy-momentum tensor, respectively. From the energy-momentum tensor equation for the $f(R,T)=R+f(T)$ case, we calculate the form of the geometric entropy in such a theory. Then, the generalized second law of thermodynamics is quantified and some relations for its obedience in $f(R,T)$ gravity are presented. Those relations depend on some cosmological quantities, as the Hubble and deceleration parameters, and also on the form of $f(T)$.

\end{abstract}


\section{Introduction}\label{sec:int}

The second law of thermodynamics states that changes of a closed thermodynamics system take place in the direction in which entropy increases. Motivated by the similarities between the physics of black holes (BHs) and thermodynamics, \cite{bekenstein/1973} has derived the generalized second law of thermodynamics (GSL), which states that the common entropy in a BH exterior plus the BH entropy never decreases. Such a generalization is supported by the information theory, as demonstrated by the author.

Since BHs are characterized by strong gravitational fields in quantum length scales, it is expected that quantum theories of gravity may bring up some new features and insights about BH physics. Indeed, this is the case for loop quantum gravity (\cite{rovelli/1990,rovelli/1998}). \cite{rovelli/1996}, for instance, has obtained a statistical entropy proportional to the area of the BH, as in the Bekenstein relation. \cite{ashtekar/1998} also have shown that the entropy of a large nonrotating BH is proportional to its horizon area. By comparison with the Bekenstein formula, \cite{meissner/2004} has fixed the value of the quantum of area.

Not only such quantum regimes stimulate the development of generalized gravity theories. Also the accelerated expansion our universe is undergoing (\cite{riess/1998,perlmutter/1999}) yields the elaboration of gravitational theories whose derived cosmological models are able to describe such a universe dynamics without the necessity of invoking the existence of exotic fluids, such as dark energy in $\Lambda$CDM model.

Among those alternatives, one could quote the $f(R)$ (\cite{sotiriou/2010,defelice/2010,capozziello/2011,nojiri/2006}) and $f(R,T)$ theories (\cite{harko/2011}). Well behaved cosmological models have been derived from such alternative theories of gravity, as one can check, for instance, in (\cite{hu/2007,navarro/2007,clifton/2015,moraes/2014b,moraes/2015,moraes/2015b,rao/2015,Jamil et ala}).

In the $f(R)$ gravity realm, \cite{eling/2006} have derived the Einstein's field equations (FEs) from the Clausius relation in thermodynamics. By assuming the geometric entropy is given by a quarter of the apparent horizon area, \cite{akbar/2006} have applied the first law of thermodynamics to the apparent horizon of a Friedmann-Robertson-Walker (FRW) universe and obtained the Friedmann equations for a flat universe. GSL has also been approached from the $f(R)$ gravity perspective (\cite{wu/2008,mohseni/2007,karami/2012,herrera/2014}).

On the other hand, in what concerns $f(R,T)$ gravity, \cite{harko/2014} presented the open irreversible thermodynamic interpretation of a cosmological model. It has also been shown that due to the matter-geometry coupling, predicted by such theories, during the cosmological evolution, a large amount of comoving entropy could be produced. \cite{houndjo/2014} have considered thermodynamics in a Little Rip cosmological scenario and showed that the second law is always satisfied for such a universe for the temperature inside the horizon being the same as that of the apparent horizon. Furthermore, in \citep{Jamil et alb} a systematic study of the first law of thermodynamics was presented. The authors have demonstrated that the first law is violated because of entropy productions in $f(R,T)$ theory\footnote{The violation of the first law of thermodynamics is not an exclusive $f(R,T)$ gravity feature. \cite{miao/2011} have proved that it also does not hold in $f(T)$ gravity, for instance, with $T$ being the torsion scalar.}. A non-equilibrium picture of thermodynamics was discussed at the apparent horizon of FRW universe by \cite{sharif/2012}. The validity of first and second laws was checked for such a scenario. 

Note that due to $f(R,T)$ gravity recent elaboration, a robust study of GSL from such a perspective still lacks. Moreover, as argued by the $f(R,T)$ gravity authors, the dependence of the gravitational part of the action on the trace of the energy-momentum tensor would come from the consideration of quantum effects (which are neglected in standard gravity and $f(
R)$ theories, for instance). Meanwhile, the corrections provided to the entropy-area relationship quoted above also have a quantum nature. For instance, the well-known power-law correction to such a relationship (given by $S_A=A/4$) reads

\begin{equation}\label{1}
S_A=\frac{A}{4}\left[1-\frac{\alpha(4\pi)^{\frac{\alpha}{2}-1}}{(4-\alpha)r_c^{2-\alpha}} A^{1-\frac{\alpha}{2}}\right],
\end{equation}
with $\alpha$ being a constant and $r_c$ a cross-over scale, and appears due to the consideration of the entanglement of quantum fields as shown by \cite{sheykhi/2011}. The logarithmic correction, given by (\cite{sadjadi/2011})

\begin{equation}\label{2}
S_A=\frac{A}{4}+\beta \log\left(\frac{A}{4}\right)+\gamma,
\end{equation}
with $\beta$ and $\gamma$ being constants, arise due to quantum fluctuations according to \cite{rovelli/1996}.

Those modifications in the entropy-area relationship yields corrections in the Einstein-Hilbert action. Such a combination motivates the study of GSL from the $f(R,T)$ gravity perspective. In this way, that is the aim of the present work, i.e., to revisit GSL in $f(R,T)$ gravity and obtain the referring expression for the geometric entropy. We must also explore GSL in the $f(R,T)$ cosmological context, since the validity of it will depend directly on cosmological parameters. Such an approach was applied in a number of other alternative cosmological models as one can check in (\cite{ford/2001,mohseni/2006,setare/2007,jamil/2010a,jamil/2010b}). In fact, GSL has been broadly investigated in alternative gravity theories (check \cite{abdolmaleki/2014,herrera/2014b,bamba/2013,ghosh/2013,chattopadhyay/2012,mazumder/2011,karami/2011,karami/2010}).

It should also be mentioned that there have already been some attempts to prove GSL in different ways, such as by quantum information theory (\cite{hosoya/2001,song/2008}) and by using adiabatically collapsing thick light shells (\cite{he/2007}). Moreover, a simple and direct proof of GSL was obtained for a quasistationary semiclassical BH in (\cite{frolov/1993}).

\section{The $f(R,T)$ theory of gravity}\label{sec:frt}

Recent elaborated by T. Harko and collaborators (\cite{harko/2011}), the $f(R,T)$ theories of gravity consider in the gravitational part of the action a general dependence of both the Ricci scalar $R$ and the trace of the energy-momentum tensor $T$. The latter appears if one, departing from the general relativity and $f(R)$ theory cases, considers the existence of quantum effects. The total action in $f(R,T)$ theories reads

\begin{equation}\label{frt1}
S=\frac{1}{16\pi}\int f(R,T)\sqrt{-g}d^4x+\int \mathcal{L}_m\sqrt{-g}d^4x,
\end{equation}
for which $f(R,T)$ is the ``general"\footnote{As we shall revisit later, this function is not indeed general or arbitrary, since it is submissive to some physical conditions.} function of $R$ and $T$, and $\mathcal{L}_m$ is the matter lagrangian.

By varying action (\ref{frt1}) with respect to the metric, one obtains

\begin{eqnarray}\label{frt2}
f_R(R,T)R_{\mu\nu}-\frac{1}{2}f(R,T)g_{\mu\nu}+(g_{\mu\nu}\Box-\nabla_\mu\nabla_\nu)\nonumber\\
f_R(R,T)=8\pi T_{\mu\nu}-f_T(R,T)T_{\mu\nu}-f_T(R,T)\Theta_{\mu\nu}.
\end{eqnarray}
In (\ref{frt2}), $R_{\mu\nu}$ represents the Ricci tensor and $\nabla_\mu$ the covariant derivative with respect to the symmetric connection associated to $g_{\mu\nu}$, $f_R(R,T)\equiv\partial f(R,T)/\partial R$, $f_T(R,T)\equiv\partial f(R,T)/\partial T$, $\Box\equiv\partial_\mu(\sqrt{-g}g^{\mu\nu}\partial_\nu)/\sqrt{-g}$, $T_{\mu\nu}=g_{\mu\nu}\mathcal{L}_m-2\partial\mathcal{L}_m/\partial g^{\mu\nu}$ is the energy-momentum tensor, which will be assumed as that of a perfect fluid and $\Theta_{\mu\nu}\equiv g^{\alpha\beta}\delta T_{\alpha\beta}/\delta g^{\mu\nu}=-2T_{\mu\nu}-pg_{\mu\nu}$, with $p$ being the pressure of the universe.

We will assume $f(R,T)=R+f(T)$ in (\ref{frt2}), with $f(T)$ being a generic function of $T$. Such a functional form for $f(R,T)$ has been broadly investigated and resulted in well-behaved cosmological models, as one can check in (\cite{harko/2011,moraes/2014b,moraes/2015,moraes/2015b,rao/2015,f1,f2,f3,f4,f5,f6}), among many others. Such a substitution yields for the energy-momentum tensor of the model, the following:

\begin{equation}\label{frt3}
T_{\mu\nu}=\frac{1}{8\pi+f_T}\left\{G_{\mu\nu}-\left[\frac{1}{2}f(T)+f_Tp\right]g_{\mu\nu}\right\},
\end{equation}
in which $G_{\mu\nu}$ is the usual Einstein tensor. Note that when $f(T)=0$ in the above equation, the usual relation for the energy-momentum tensor of General Relativity is recovered.

To have a complete and fully satisfactory description of thermodynamics in $f(R,T)$ gravity, we need to know what is the form of the geometric entropy in this type of modified gravity. We will address this question in the next section.

\section{Derivation of geometric entropy in $f(R,T)$ theory via Wald's approach}
It was motivated by Wald to introduce geometric entropy for any type of second order action of modified gravity by starting from the first law 
of thermodynamics \citep{wald}. To compute entropy using this elegant method, we should consider 
 the net heat flux which is passing through an open patch
$d\mathcal{H}=dAd\lambda$, on a null surface of the BH horizon, i.e., to evaluate the following expression:
\begin{equation}
\delta Q=\int\limits_\mathcal{H} T_{\mu\nu}\xi^\mu k^\nu dAd\lambda.\label{Q}
\end{equation}
In Eq.(\ref{Q}), $Q$ is defined as the net heat flux, $T_{\mu\nu}$ as the effective energy-momentum tensor, $\xi^\mu$ is the killing horizon vector, $\mathcal{H}$ denotes the null surface, $\lambda$ is an appropriate affine parameter to parametrize the geodesics and the vector $k^\mu=\frac{dx^\mu}{d\lambda}$ is the tangent vector to $\mathcal{H}$.


To evaluate (\ref{Q}) we need to specify the model of gravity. In the present case, we must use $T_{\mu\nu}$ from $f(R,T)$ gravity, i.e., Eq.(\ref{frt3}). By doing this we obtain:

\begin{eqnarray}\label{s1}
\delta Q&=&\int\limits_\mathcal{H} \left\{ R_{\mu\nu}-\frac{1}{2}g_{\mu\nu}[R+f(T)+2f_Tp]\right\}\frac{\xi^\mu k^\nu dAd\lambda}{8\pi+f_T},\nonumber\\
&=&\int\limits_\mathcal{H} \left( \nabla_\mu\nabla_\nu \xi^\mu-\frac{1}{2}Pg_{\mu\nu}\xi^{\mu}\right)\frac{ k^\nu}{8\pi+f_T} dAd\lambda.
\end{eqnarray}
In (\ref{s1}), $P\equiv R+f(T)+f_Tp$ and we have used the identity $R_{\mu\nu}\xi^\mu=\nabla_\mu\nabla_\nu \xi^\mu$,  which is valid only for an exact Killing vector
$\xi^\mu$. 

It is possible to make more simplifications using integration of (\ref{s1}) part-by-part as the following (note that the second term in (\ref{s1}) vanishes because $\xi^{\nu}k_{\nu}=0$ \citep{horizon}):
\begin{eqnarray}\label{deltaQ-n}
\delta Q&=&\int\limits_\mathcal{H} k^\nu\nabla_\mu\Big[\nabla_\nu\Big( \frac{\xi^\mu}{8\pi+f_T} \Big) \Big]dAd\lambda,\nonumber\\
&=& \int\limits_\mathcal{H} k^\nu l^\mu \nabla_\nu\Big( \frac{\xi_\mu}{8\pi+f_T} \Big) dAd\lambda,\nonumber\\
&=& \int\limits_\mathcal{H} k^\nu l^\mu \Big[ \frac{\nabla_\nu\xi_\mu}{8\pi+f_T} - \frac{\xi_\mu\nabla_\nu f_T}{(8\pi+f_T)^2} \Big] dAd\lambda, \nonumber\\
&=&\frac{1}{8\pi}\int\limits_\mathcal{H} k^\nu l^\mu\Big[
\nabla_\nu\xi_\mu \sum_{n=0}^{\infty}\Big(-\frac{f_T}{8\pi}\Big)^n\\&&\nonumber - \frac{\xi_\mu\nabla_\nu f_T }{8\pi}\sum_{n=0}^{\infty}(n+1)\Big(-\frac{f_T}{8\pi}\Big)^n\Big]
dAd\lambda.\nonumber
\end{eqnarray}
So far, the above expression gives us the functional form of heat and we need to have an explicit non-local form for it. Regarding this item, we assume that 
the first order derivative of $f(T)$ with respect to $T$ is small, i.e., $f_T\ll1$. So, we can expand (\ref{deltaQ-n}) in Taylor's series as the following:
\begin{eqnarray}
\delta Q&=&\frac{1}{8\pi}\int\limits_\mathcal{H} k^\nu l^\mu \nabla_\nu \xi_\mu dAd\lambda\\&&\nonumber-\frac{1}{(8\pi)^2}\int\limits_\mathcal{H} k^\nu l^\mu f_T\nabla_\nu \xi_\mu dAd\lambda\\&&\nonumber\\&&\nonumber+\frac{1}{(8\pi)^3}\int\limits_\mathcal{H} k^\nu l^\mu f_T^2\nabla_\nu \xi_\mu dAd\lambda-\ldots \nonumber\\&&
-\frac{1}{8\pi}\int\limits_\mathcal{H} k^\nu l^\mu \xi_\mu\nabla_\nu f_T  dAd\lambda\\&&\nonumber+
\frac{2}{(8\pi)^2}\int\limits_\mathcal{H} k^\nu l^\mu f_T\xi_\mu\nabla_\nu f_T  dAd\lambda
\\&&\nonumber-\frac{3}{(8\pi)^3}\int\limits_\mathcal{H} k^\nu l^\mu f_T^2\xi_\mu\nabla_\nu f_T  dAd\lambda \nonumber\\&&+\ldots \nonumber
\end{eqnarray}

\begin{eqnarray}\label{deltaQ1}
\delta Q&=&\frac{\kappa}{2\pi}\left.\Big(\frac{dA}{4}\Big)\right\vert
_{0}^{d\lambda}-\frac{1}{8\pi}\frac{\kappa}{2\pi}\left.\Big(\frac{f_TdA}{4}\Big)\right\vert
_{0}^{d\lambda}\\&&\nonumber+\frac{1}{64\pi^2}\frac{\kappa}{2\pi}\left.\Big(\frac{f_T^2dA}{4}\Big)\right\vert
_{0}^{d\lambda}-\ldots
\\&&\nonumber
-\frac{1}{8\pi}\int\limits_\mathcal{H} k^\nu l^\mu \xi_\mu\nabla_\nu f_T  dAd\lambda\\&&\nonumber+
\frac{2}{(8\pi)^2}\int\limits_\mathcal{H} k^\nu l^\mu f_T\xi_\mu\nabla_\nu f_T  dAd\lambda
\\&&\nonumber-\frac{3}{(8\pi)^3}\int\limits_\mathcal{H} k^\nu l^\mu f_T^2\xi_\mu\nabla_\nu f_T  dAd\lambda \nonumber\\&&+\ldots, \nonumber
\end{eqnarray}
with $\mathcal{T}=\kappa/2\pi$ and $\mathcal{T}$ being the temperature.

Due to the extra non-local dissipative term in (\ref{deltaQ1}), we observe that 
the 
first law of BH thermodynamics is violated for $f(R,T)=R+f(T)$
gravity.
We can rewrite 
(\ref{deltaQ1}) in terms of $\mathcal{T}$ and entropy corrections $S_i$ as follows:
\begin{eqnarray}\label{deltaQ}
\delta Q&=&\mathcal{T}\delta S-\mathcal{T}\delta S_1 +\mathcal{T}\delta S_2-\ldots
\\&&\nonumber-\frac{1}{8\pi}\int\limits_\mathcal{H} k^\nu l^\mu \xi_\mu\nabla_\nu f_T  dAd\lambda+\ldots
\end{eqnarray}
where
\begin{equation}
\delta S_1= \frac{1}{8\pi}\left.\Big(\frac{dA}{4}\Big)\right\vert
_{0}^{d\lambda},\ \ \delta S_2=\frac{1}{(8\pi)^2}\left.\Big(\frac{f_TdA}{4}\Big)\right\vert
_{0}^{d\lambda},\ldots\nonumber
\end{equation}

Finally we can present the ultimate  form of geometric entropy in $f(R,T)$ model  as follows:
\begin{eqnarray}\label{entropy}
&&S=\frac{A}{4}-\frac{1}{8\pi}\frac{f_TA}{4}+\frac{1}{64\pi^2}\frac{f_T^2A}{4}\\&&\nonumber -\frac{1}{4\kappa}\int\limits_\mathcal{H} k^\nu l^\mu\xi_\mu\nabla_\nu f_T  dAd\lambda-\ldots
\end{eqnarray}
As we mentioned before, the last terms indicate an entropy production in $f(R,T)$ theory. These terms exist even if the background metric is time-independent.

We can rewrite (\ref{entropy}) in a much more elegant form by using the expression of entropy in Einstein gravity. As we know, the expression of Bekenstein-Hawking entropy is given by 
 $S_{BH}\equiv\frac{A}{4}$ \citep{Jacobson et al. 1994} (for $f(R)$ gravity, see \citep{ Cognola et al. 2005},\citep{Bamba and Geng 2009},\citep{Bamba and Geng  2010}). So, we can rewrite (\ref{entropy}) in the following equivalence form:
\begin{equation}\label{entropy2}
S=S_{BH}\Big[1-\frac{f_T}{8\pi}+\frac{f_T^2}{(8\pi)^2} -\ldots\Big]\equiv \frac{A}{4}\frac{1}{1+\frac{f_T}{8\pi}}.
\end{equation}
We indicate that the first term in (\ref{entropy2}) 
is the Wald's (Bekenstein-Hawking)
entropy for Einstein gravity which can be obtained by several methods. The other terms 
arise due to the coupling of curvature with matter in the $f(R,T)$ gravity action. 
\section{Generalized second law of thermodynamics}  
Let us quantify GSL for a BH in $f(R,T)$ gravity. The GSL states that thermodynamic entropy of a BH added with the entropy of the cosmological background must be a monotonic-increasing function of time. Considering $S_i$ as the amount of entropy of the matter components, from the first law of thermodynamics we know that the infinitesimal difference of the entropy, energy density and  volume are related by the following equation:
\begin{equation}
\mathcal{T}_idS_i=d(\rho_iV)+p_idV-\mathcal{T}_idS_p\,, \label{6}
\end{equation}
where $\rho_i$ is the density and $S_p$ is the entropy production. Furthermore, the enclosed volume of the horizon reads $V=\frac{4\pi}{3}R_\mathcal{H}^3$ . 

We remind that the continuity equation is the following:

\begin{equation} \label{7}
\dot\rho_i+3H(\rho_i+p_i)=Q_i\,, 
\end{equation} 
for which $H=\frac{\dot{a}}{a}$ is the Hubble parameter and $a$ is the scale factor. 

We suppose that different components are interacting with each other through a set of interaction terms named $Q_i$. Generally speaking, these interaction terms collectively satisfy $\sum_i Q_i=0$. 
From Eqs.(\ref{6}) and (\ref{7}), we get
\begin{equation}\label{8}
\dot S_i=\frac{4\pi}{3}R_\mathcal{H}^3\frac{Q_i}{\mathcal{T}_i}+4\pi R_\mathcal{H}^2(\dot R_\mathcal{H}-HR_\mathcal{H})\left(\frac{\rho_i+p_i}{\mathcal{T}_i}\right)-\dot S_p\,.
\end{equation} 
We need to suppose that there is a thermal equilibrium between the fluids and the horizon. So, we assume that
$\mathcal{T}_i=\mathcal{T}_\mathcal{H}$. Consequently 
\begin{equation}\label{8a}
\dot S_I+\dot S_p=-\frac{8\pi \dot{H}R_\mathcal{H}^2(\dot{R}_\mathcal{H}-HR_\mathcal{H})}{\mathcal{T}_\mathcal{H}},
\end{equation}
here we define the total internal entropy of the fluid by $S_I=\sum_i S_i$. The form of GSL reads 
\begin{equation}\label{x1}
\dot S_{tot}=\dot S_I+\dot S_p+\dot S\geq0\,,
\end{equation}
where $\dot S$ presumably is the time derivative of (\ref{entropy2}), which is given by:
 \begin{equation}\label{8a}
\dot S=S\Big(2\frac{\dot{R}_\mathcal{H}}{R_\mathcal{H}}+\frac{\dot{F}}{F}\Big),
\end{equation}
with

\begin{equation}
F\equiv \frac{1}{1+\frac{f_T}{8\pi}},
\end{equation}

\begin{equation}\label{F}
\frac{\dot{F}}{F}=-\frac{\dot{T}f_{TT}}{8\pi+f_T}.
\end{equation}

So, what we need is to check where the following expression is respected
\begin{equation}\label{GSL}
2\frac{\dot{R}_\mathcal{H}}{R_\mathcal{H}}+\frac{\dot{F}}{F}> \frac{8\dot{H}(\dot{R}_\mathcal{H}-HR_\mathcal{H})}{\mathcal{T}_\mathcal{H}},
\end{equation}
since $\dot{S}_{tot}$ must be $>0$.

\subsection{
GSL on dynamical apparent horizon}
It was demonstrated that in an accelerating universe, the GSL holds only in the case where  
the boundary surface is the apparent horizon. Using the event horizon we cannot obtain an accelerating scenario \citep{zhou}. So, we conclude that the event horizon is not a physical boundary from the point of view of thermodynamics. Because of this, we
consider the dynamical apparent horizon \citep{Cai and Kim 2005}
\begin{equation}\label{14}
R_A=\frac{1}{H}\,.
\end{equation}
The apparent horizon $R_A$ is a marginally trapped surface with vanishing expansion and is determined from the condition $g^{ij}\partial_i\tilde{r}\partial_j\tilde{r}=0$, where $\tilde{r}=r(t)a(t)$ and $i,j =0,1$ \citep{Hayward 1998}, \citep{Cai and Cao 2007}. Assuming $A = 4\pi R_A^2$ , we rewrite (\ref{GSL}) using $R_H\equiv R_A$:

\begin{eqnarray}\label{app1}
&&\dot S_{tot}= \frac{16\pi^2\dot{H}}{bH^3}\Big(1+\frac{\dot{H}}{H^2}\Big)+\frac{\pi }{H^2}\Big(2\frac{\dot{H}}{H}+\frac{\dot{F}}{F}\Big)> 0.
\end{eqnarray}
An equivalent form of (\ref{app1}) is given by:

\begin{eqnarray}\label{app11}
&&\dot S_{tot}= \frac{16\pi^2q(q+1)}{bH}+\frac{\pi }{H^2}\Big(2\frac{\dot{H}}{H}+\frac{\dot{F}}{F}\Big)>0,
\end{eqnarray}
with $q=-\Big(1+\frac{\dot{H}}{H^2}\Big)$ being the deceleration parameter. Also, we have used the thermal equilibrium for our thermodynamical system and $\mathcal{T}_H =\frac{ bH}{2\pi}$, with $b$ being a constant
\citep{Akbar 2009}. 
 
Since the present value of $q$ is $\sim-0.5$ from recent cosmic microwave background observations (\cite{planckcollaboration/2013,hinshaw/2013}), the following relation must be satisfied in order to GSL be respected:

\begin{equation}\label{xx1}
\frac{\dot{F}}{F}>H\left(\frac{4\pi}{b}+1\right),
\end{equation}
with the present value of $H$ being $H_0=70km/s/Mpc$ (\cite{hinshaw/2013}).

\subsection{GSL on event horizon}
The future event horizon is defined as the distance that light travels from the present time to infinity and is defined as:

\begin{equation}\label{RE}
R_E=a(t)\int_{t}^{\infty}\frac{dt'}{a(t')},
\end{equation}

It is straighforward to prove that $\dot{R}_E = H R_E - 1$. We rewrite (\ref{GSL}) using (\ref{RE}) and obtain:
\begin{eqnarray}\label{eve1}
&&\dot S_{tot}=\frac{16\pi^2 \dot{H}R_E^2}{bH}+\pi R_E^2\Big(2\frac{\dot{R}_E}{R_E}+\frac{\dot{F}}{F}\Big)> 0.
\end{eqnarray}
By using the cosmic microwave background data once again, we obtain that

\begin{equation}\label{xx2}
\frac{\dot{F}}{F}>2\left(\frac{4\pi}{b}H-\frac{\dot{R}_E}{R_E}\right).
\end{equation}

\section{Discussion}

The $f(R,T)$ theory of gravity was recently elaborated by T. Harko and collaborators (\cite{harko/2011}). Despite its recent elaboration, it already presents plenty of cosmological and astrophysical applications (in addition to the references in Introduction, one could check (\cite{ms/2016,mam/2015})).

Although $f(R,T)$ gravity also has been investigated from a thermodynamics approach (check references in Introduction), the GSL still had to be carefully studied in such a theory.

That was our aim in this article. From the energy-momentum tensor in $f(R,T)$ gravity (\ref{frt3}), we have obtained the geometric entropy for such a theory (\ref{entropy2}), which can retrieve the standard result in a certain regime. Then, we finally quantified GSL in $f(R,T)$ gravity. It was developed for both dynamical apparent and event horizons. In both cases, relations for the obedience of GSL as functions of the specific functional form adopted for $f(T)$ and cosmological parameters, such as $a$, $H$ and $q$, were found.  

It is not a novelty the combination of thermodynamics concepts with cosmology. \cite{padmanabhan/2005} has discovered some laws that connect thermodynamics with Einstein's field equations. Moreover it was shown that the Friedmann equations can be obtained from the application of the first law of thermodynamics to the apparent horizon of an FRW universe when one assumes the geometric entropy is given by a quarter of the apparent horizon area (\cite{Cai and Kim 2005}).

An important consequence of Eqs.(\ref{xx1}) and (\ref{xx2}) is that we can obtain from them some constraints to be put in the functional form of $f(T)$. Although normally it is said that $f(R,T)$ is a general function of $R$ and $T$, by carefully investigating the $f(R,T)$ gravity, one realizes that it is not.

From the analysis of $f(R,T)$ cosmological models in phase space, the functionality of $f(T)$ was confined to a particular form in (\cite{models}). From the adiabatic condition, \cite{sun/2015} have found out a constraint relation between $f(T)$ and the equation of state of the universe. The coincidence problem (\cite{zlatev/1999}) was addressed by \cite{rudra/2015}, which leaded to a filtration of various models of $f(R,T)$ gravity. 

By recalling Eq.(\ref{F}) together with (\ref{xx1}) and (\ref{xx2}), it is clear that GSL can provide more constraints to the choice of a specific form for $f(T)$ in $f(R,T)$ theory. This departs from the GSL analysis from the teleparallel gravity perspective (\cite{bamba/2013}), in which the GSL obedience relations weakly depend on the choice of $f(T)$ (remind that in this case, $T$ is the scalar torsion).

The {\it lhs} of both Eqs.(\ref{xx1}) and (\ref{xx2}) depends on the time derivative of the trace of the energy-momentum tensor. From the equation for the energy-momentum tensor shown in Section \ref{sec:frt}, it is intuitive that its time derivative shall be negative (some plots of the evolution of the trace of the energy-momentum tensor in $f(R,T)$ gravity which support this argument can be checked in (\cite{ms/2016})). This feature allied with the minus sign in the {\it lhs} of Eq.(\ref{xx1}) guarantees that the {\it lhs} is positive (as it must be) for $f_{TT}>0$ and $f_T>-8\pi$.

By carefully analysing the term $4\pi H/b-\dot{R}_E/R_E$, the same kind of constraints can be obtained from Eq.(\ref{xx2}). In this way, the apparent arbitrariness of $f(T)$ fade away and allied with some other studies, such as cosmology in phase space or adiabatic condition, the form of $f(T)$ in $f(R,T)$ gravity shall be fine tuned, yielding more reliable and realistic $f(R,T)$ models.\\

{\bf Acknowledgements}

PHRSM would like to thank Sao Paulo Research Foundation (FAPESP), grant 2015/08476-0, for financial support.

\bibliographystyle{spr-mp-nameyear-cnd}  

\end{document}